\documentclass[prl,superscriptaddress,twocolumn,showpacs,amsmath,amssymb]{revtex4}

\usepackage{epsfig}
\usepackage{dcolumn}   
\usepackage{bm}        
\usepackage{ulem}

\usepackage{graphicx}  

\def\met{\mbox{$\not\!\!{E}_{T}\,$}}

\def\Zll{\mbox{$Z\rightarrow \ell \ell$}}
\def\Znn{\mbox{$Z\rightarrow\nu\overline{\nu}$}}
\def\Zllp{\mbox{$Z(\rightarrow \ell \ell)$}}

\def\Wen{\mbox{$W\rightarrow e\nu$}}
\def\Wmn{\mbox{$W\rightarrow\mu\nu$}}
\def\Wtn{\mbox{$W\rightarrow\tau\nu$}}
\def\Wln{\mbox{$W\rightarrow\ell\nu$}}
\def\Wlnp{\mbox{$W(\rightarrow\ell\nu)$}}
\def\Wenp{\mbox{$W(\rightarrow e\nu)$}}
\def\Wmnp{\mbox{$W(\rightarrow\mu\nu)$}}

\def\pyt{{\sc pythia}}
\def\geant3{{\sc geant3}}

\begin{document}

\title{\bf Search for New Physics with a Dijet plus Missing $E_T$ Signature in $p\bar{p}$ Collisions at $\sqrt{s}=$~1.96~TeV}

\date{\today}

\affiliation{Institute of Physics, Academia Sinica, Taipei, Taiwan 11529, Republic of China} 
\affiliation{Argonne National Laboratory, Argonne, Illinois 60439} 
\affiliation{University of Athens, 157 71 Athens, Greece} 
\affiliation{Institut de Fisica d'Altes Energies, Universitat Autonoma de Barcelona, E-08193, Bellaterra (Barcelona), Spain} 
\affiliation{Baylor University, Waco, Texas  76798} 
\affiliation{Istituto Nazionale di Fisica Nucleare Bologna, $^{cc}$University of Bologna, I-40127 Bologna, Italy} 
\affiliation{Brandeis University, Waltham, Massachusetts 02254} 
\affiliation{University of California, Davis, Davis, California  95616} 
\affiliation{University of California, Los Angeles, Los Angeles, California  90024} 
\affiliation{University of California, San Diego, La Jolla, California  92093} 
\affiliation{University of California, Santa Barbara, Santa Barbara, California 93106} 
\affiliation{Instituto de Fisica de Cantabria, CSIC-University of Cantabria, 39005 Santander, Spain} 
\affiliation{Carnegie Mellon University, Pittsburgh, PA  15213} 
\affiliation{Enrico Fermi Institute, University of Chicago, Chicago, Illinois 60637}
\affiliation{Comenius University, 842 48 Bratislava, Slovakia; Institute of Experimental Physics, 040 01 Kosice, Slovakia} 
\affiliation{Joint Institute for Nuclear Research, RU-141980 Dubna, Russia} 
\affiliation{Duke University, Durham, North Carolina  27708} 
\affiliation{Fermi National Accelerator Laboratory, Batavia, Illinois 60510} 
\affiliation{University of Florida, Gainesville, Florida  32611} 
\affiliation{Laboratori Nazionali di Frascati, Istituto Nazionale di Fisica Nucleare, I-00044 Frascati, Italy} 
\affiliation{University of Geneva, CH-1211 Geneva 4, Switzerland} 
\affiliation{Glasgow University, Glasgow G12 8QQ, United Kingdom} 
\affiliation{Harvard University, Cambridge, Massachusetts 02138} 
\affiliation{Division of High Energy Physics, Department of Physics, University of Helsinki and Helsinki Institute of Physics, FIN-00014, Helsinki, Finland} 
\affiliation{University of Illinois, Urbana, Illinois 61801} 
\affiliation{The Johns Hopkins University, Baltimore, Maryland 21218} 
\affiliation{Institut f\"{u}r Experimentelle Kernphysik, Karlsruhe Institute of Technology, D-76131 Karlsruhe, Germany} 
\affiliation{Center for High Energy Physics: Kyungpook National University, Daegu 702-701, Korea; Seoul National University, Seoul 151-742, Korea; Sungkyunkwan University, Suwon 440-746, Korea; Korea Institute of Science and Technology Information, Daejeon 305-806, Korea; Chonnam National University, Gwangju 500-757, Korea; Chonbuk National University, Jeonju 561-756, Korea} 
\affiliation{Ernest Orlando Lawrence Berkeley National Laboratory, Berkeley, California 94720} 
\affiliation{University of Liverpool, Liverpool L69 7ZE, United Kingdom} 
\affiliation{University College London, London WC1E 6BT, United Kingdom} 
\affiliation{Centro de Investigaciones Energeticas Medioambientales y Tecnologicas, E-28040 Madrid, Spain} 
\affiliation{Massachusetts Institute of Technology, Cambridge, Massachusetts  02139} 
\affiliation{Institute of Particle Physics: McGill University, Montr\'{e}al, Qu\'{e}bec, Canada H3A~2T8; Simon Fraser University, Burnaby, British Columbia, Canada V5A~1S6; University of Toronto, Toronto, Ontario, Canada M5S~1A7; and TRIUMF, Vancouver, British Columbia, Canada V6T~2A3} 
\affiliation{University of Michigan, Ann Arbor, Michigan 48109} 
\affiliation{Michigan State University, East Lansing, Michigan  48824}
\affiliation{Institution for Theoretical and Experimental Physics, ITEP, Moscow 117259, Russia} 
\affiliation{University of New Mexico, Albuquerque, New Mexico 87131} 
\affiliation{Northwestern University, Evanston, Illinois  60208} 
\affiliation{The Ohio State University, Columbus, Ohio  43210} 
\affiliation{Okayama University, Okayama 700-8530, Japan} 
\affiliation{Osaka City University, Osaka 588, Japan} 
\affiliation{University of Oxford, Oxford OX1 3RH, United Kingdom} 
\affiliation{Istituto Nazionale di Fisica Nucleare, Sezione di Padova-Trento, $^{dd}$University of Padova, I-35131 Padova, Italy} 
\affiliation{LPNHE, Universite Pierre et Marie Curie/IN2P3-CNRS, UMR7585, Paris, F-75252 France} 
\affiliation{University of Pennsylvania, Philadelphia, Pennsylvania 19104}
\affiliation{Istituto Nazionale di Fisica Nucleare Pisa, $^{ee}$University of Pisa, $^{ff}$University of Siena and $^{gg}$Scuola Normale Superiore, I-56127 Pisa, Italy} 
\affiliation{University of Pittsburgh, Pittsburgh, Pennsylvania 15260} 
\affiliation{Purdue University, West Lafayette, Indiana 47907} 
\affiliation{University of Rochester, Rochester, New York 14627} 
\affiliation{The Rockefeller University, New York, New York 10021} 
\affiliation{Istituto Nazionale di Fisica Nucleare, Sezione di Roma 1, $^{hh}$Sapienza Universit\`{a} di Roma, I-00185 Roma, Italy} 

\affiliation{Rutgers University, Piscataway, New Jersey 08855} 
\affiliation{Texas A\&M University, College Station, Texas 77843} 
\affiliation{Istituto Nazionale di Fisica Nucleare Trieste/Udine, I-34100 Trieste, $^{ii}$University of Trieste/Udine, I-33100 Udine, Italy} 
\affiliation{University of Tsukuba, Tsukuba, Ibaraki 305, Japan} 
\affiliation{Tufts University, Medford, Massachusetts 02155} 
\affiliation{Waseda University, Tokyo 169, Japan} 
\affiliation{Wayne State University, Detroit, Michigan  48201} 
\affiliation{University of Wisconsin, Madison, Wisconsin 53706} 
\affiliation{Yale University, New Haven, Connecticut 06520} 
\author{T.~Aaltonen}
\affiliation{Division of High Energy Physics, Department of Physics, University of Helsinki and Helsinki Institute of Physics, FIN-00014, Helsinki, Finland}
\author{J.~Adelman}
\affiliation{Enrico Fermi Institute, University of Chicago, Chicago, Illinois 60637}
\author{B.~\'{A}lvarez~Gonz\'{a}lez$^v$}
\affiliation{Instituto de Fisica de Cantabria, CSIC-University of Cantabria, 39005 Santander, Spain}
\author{S.~Amerio$^{dd}$}
\affiliation{Istituto Nazionale di Fisica Nucleare, Sezione di Padova-Trento, $^{dd}$University of Padova, I-35131 Padova, Italy} 

\author{D.~Amidei}
\affiliation{University of Michigan, Ann Arbor, Michigan 48109}
\author{A.~Anastassov}
\affiliation{Northwestern University, Evanston, Illinois  60208}
\author{A.~Annovi}
\affiliation{Laboratori Nazionali di Frascati, Istituto Nazionale di Fisica Nucleare, I-00044 Frascati, Italy}
\author{J.~Antos}
\affiliation{Comenius University, 842 48 Bratislava, Slovakia; Institute of Experimental Physics, 040 01 Kosice, Slovakia}
\author{G.~Apollinari}
\affiliation{Fermi National Accelerator Laboratory, Batavia, Illinois 60510}
\author{A.~Apresyan}
\affiliation{Purdue University, West Lafayette, Indiana 47907}
% ...

\collaboration{CDF Collaboration\footnote{With visitors from $^a$University of Massachusetts Amherst, Amherst, Massachusetts 01003,
$^b$Universiteit Antwerpen, B-2610 Antwerp, Belgium, 
$^c$University of Bristol, Bristol BS8 1TL, United Kingdom,
$^d$Chinese Academy of Sciences, Beijing 100864, China, 
$^e$Istituto Nazionale di Fisica Nucleare, Sezione di Cagliari, 09042 Monserrato (Cagliari), Italy,
$^f$University of California Irvine, Irvine, CA  92697, 
$^g$University of California Santa Cruz, Santa Cruz, CA  95064, 
$^h$Cornell University, Ithaca, NY  14853, 
$^i$University of Cyprus, Nicosia CY-1678, Cyprus, 
$^j$University College Dublin, Dublin 4, Ireland,
$^k$University of Edinburgh, Edinburgh EH9 3JZ, United Kingdom, 
$^l$University of Fukui, Fukui City, Fukui Prefecture, Japan 910-0017
$^m$Kinki University, Higashi-Osaka City, Japan 577-8502
$^n$Universidad Iberoamericana, Mexico D.F., Mexico,
$^o$University of Iowa, Iowa City, IA  52242,
$^p$Kansas State University, Manhattan, KS 66506
$^q$Queen Mary, University of London, London, E1 4NS, England,
$^r$University of Manchester, Manchester M13 9PL, England,
$^s$Muons, Inc., Batavia, IL 60510, 
$^t$Nagasaki Institute of Applied Science, Nagasaki, Japan, 
$^u$University of Notre Dame, Notre Dame, IN 46556,
$^v$University de Oviedo, E-33007 Oviedo, Spain, 
$^w$Texas Tech University, Lubbock, TX  79609, 
$^x$IFIC(CSIC-Universitat de Valencia), 56071 Valencia, Spain,
$^y$Universidad Tecnica Federico Santa Maria, 110v Valparaiso, Chile,
$^z$University of Virginia, Charlottesville, VA  22906
$^{aa}$Bergische Universit\"at Wuppertal, 42097 Wuppertal, Germany,
$^{bb}$Yarmouk University, Irbid 211-63, Jordan
$^{jj}$On leave from J.~Stefan Institute, Ljubljana, Slovenia, 
}}
\noaffiliation

\begin{abstract}
We present results of a signature-based search for new physics using a dijet 
plus missing transverse energy (\met) data sample collected in 2~fb$^{-1}$ 
of $p\bar{p}$~collisions at $\sqrt{s}$ = 1.96 TeV with the CDF~II detector at 
the Fermilab Tevatron.  We observe no significant event excess with respect to 
the standard model prediction and extract a 95\% C.L. upper limit on the cross 
section times acceptance for a potential contribution from a non-standard 
model process.  Based on this limit the mass of a potential first or second 
generation scalar leptoquark is constrained to be above 187~GeV/$c^2$.  
\end{abstract}


\maketitle

\normalsize

Events featuring two energetic jets and significant 
missing transverse energy (\met)~\cite{metdef} are a 
potential signature for phenomena not included in the 
standard model (SM), such as supersymmetry~\cite{susy},  
universal extra dimensions~\cite{ued}, and leptoquark
production~\cite{lq}.  In general, any model predicting 
pair production of unstable particles whose decay 
products are a single parton and a non-interacting 
particle could be observable as an event excess above 
the SM expectation in the dijet $+$ \met channel.  

In this Letter we report on a signature-based search for 
new physics contributions to the dijet $+$ \met final 
state in CDF Run~II data collected in $p\overline{p}$ 
collisions at $\sqrt{s}=$~1.96~TeV corresponding to an 
integrated luminosity of 2~fb$^{-1}$.  In contrast with 
previous CDF~\cite{cdflq} and D0~\cite{d0lq} searches in 
this final state, no {\it a priori} optimization of the 
kinematic selection criteria is performed to maximize 
sensitivity to a particular model.  Here the criteria are 
chosen to encompass the widest possible kinematic range 
consistent with the trigger used to collect the data 
sample.  We perform a simple counting experiment on this 
inclusive sample, comparing the number of observed events 
against the SM expectation, to search for potential 
indications of non-SM contributions.  A second, analogous 
counting experiment is made on a subsample of the highest
energy events from within the inclusive sample, which is 
a more sensitive testbed for observing contributions from 
some classes of non-SM production processes.  The tighter 
kinematic selections that define this event subset 
are chosen to give a fixed (15\%) uncertainty on the 
data-driven SM background prediction made for this sample.  
From here forward, we refer to these sets of candidate 
events as our loose and tight samples.  Based on the 
counting experiment results, we place 95\% C.L. upper 
limits on the cross section times acceptance ($\sigma 
\times A$) for a generic, non-SM process that can 
contribute events to the candidate samples.  Finally, 
we use the generic limit on $\sigma \times A$ to extract 
a lower limit on leptoquark mass for the specific case 
of scalar leptoquark production, which serves as a 
sensitivity benchmark for the result.  

A detailed description of the CDF~II detector can be found in 
Ref.~\cite{WZpaper}.  The data sample was collected using a 
three-level trigger system based on a minimum \met requirement 
of 45~GeV.  Reconstructed candidate events are required to have 
\met~$>$~80~GeV to ensure full trigger efficiency.  Jets are 
reconstructed from energy deposits in the calorimeter using a 
cone-based algorithm with a fixed radius of 0.7 in $\eta-\phi$ 
space.  The measured jet $E_{T}$ is corrected for detector  
effects and contributions from multiple $p\overline{p}$ 
interactions per bunch crossing~\cite{jetcor}.  Events in the 
candidate samples are required to have two reconstructed jets
with $|\eta|$~$<$~2.4 and $E_T$~$>$~30~GeV and no additional 
jets with $|\eta|$~$<$~3.6 and $E_T$~$>$~15~GeV.  In 
addition, the scalar sum of the two jet transverse energies, 
$H_T = E_T(\mathrm{jet1}) + E_T(\mathrm{jet2})$, must be 
greater than 125~GeV.  A separation of at least 0.5~radians 
in azimuthal angle is required between the \met and both jets 
to help suppress multi-jet backgrounds in which significant 
\met is produced by poorly measured jets.  Events from
beam-related backgrounds and cosmic rays are removed using 
standard criteria~\cite{ledprl} to tag reconstructed tracks 
and jets that are inconsistent with having been produced by
particles originating from the $p\overline{p}$ collision.   
The subset of events that satisfy tighter kinematic thresholds 
of \met~$>$~100~GeV and $H_T$~$>$~225~GeV define the tight 
candidate sample. 

A number of SM processes capable of producing the high \met 
signature in our detector contribute events to our candidate 
samples.  The largest SM background is $Z$+jets where the $Z$ 
boson decays into a pair of neutrinos.  This process results 
in a signature indistinguishable from that of potential 
signal and its relative contribution to the candidate samples 
is therefore irreducible.  The next most significant SM 
contribution is from $W$+jets in which the $W$ decays via a 
charged lepton ($e$, $\mu$, or $\tau$) and neutrino.  We 
suppress this background by rejecting events that contain 
either an isolated track~\cite{ledprl} with $p_T$~$>$~10~GeV/$c$ 
($\mu$ or $\tau$ candidate) or a jet with $E_{T}$~$>$~15~GeV 
and electromagnetic energy fraction above 90\% ($e$ candidate).

The $W/Z$+jets backgrounds are modeled using separate data 
samples collected with single lepton triggers to circumvent  
significant systematic uncertainties inherent in the 
simulation of these processes.  We estimate the number of 
SM background events from $W/Z$+jets production in our dijet 
$+$ \met candidate samples using cross section measurements  
obtained from $\Zllp$+jets and $\Wlnp$+jets ($\ell$ = $e$ 
or $\mu$) events with fully reconstructed leptons.  The 
measured cross sections contain contributions from diboson 
production where two jets are produced in the hadronic decay 
of one the bosons, and potential diboson contributions to 
the dijet $+$ \met samples are therefore included within the 
resulting background estimates.  Events in the samples used 
to make these measurements are required to have at least one 
electron ($E_T$~$>$~25~GeV) or one muon ($p_T$~$>$~20~GeV/$c$) 
passing standard selection criteria~\cite{WZpaper}.  We 
select $\Wln$ candidates by requiring \met~$>$~25~GeV for 
electrons (\met~$>$~20~GeV for muons) and $\Zll$ candidates by 
requiring a second lepton satisfying a looser set of selection 
criteria~\cite{WZpaper}.  We then apply the full set of dijet 
$+$ \met selections described previously to the selected $W/Z$ 
candidates to obtain $\Wlnp$+jets and $\Zllp$+jets event samples.  
To be consistent with the criteria used in selecting dijet $+$ 
\met signal events, reconstructed tracks and calorimeter energy 
deposits associated with the charged lepton(s) are removed prior
to application of the isolated track veto and \met requirement.  

To extract $W/Z$+jets cross sections from these samples, we 
correct for the acceptance of the $\Wln$ (25-32\%) or $\Zll$ 
(15-33\%) pieces of the selection criteria using simulated 
{\sc alpgen}~\cite{alpgen} events run through a full detector 
simulation based on \geant3~\cite{geant}.  Acceptances depend 
on the specific lepton ($\ell$ = $e$ or $\mu$) decay channel 
and on the associated loose or tight dijet $+$ \met selection 
criteria.  To account for observed differences in lepton 
reconstruction and identification efficiencies between data 
and Monte Carlo, corrections of up to 10\% per lepton are 
applied to the simulated acceptances.  Uncertainties on these 
acceptance and efficiency corrections are small ($\sim$~1-2\%) 
compared with those coming from candidate sample statistics and 
the methods used to estimate sample background contributions.  
The observed agreement in the cross section measurements made 
using high-statistics $\Wenp$+jets and $\Wmnp$+jets candidate 
samples provides validation of the techniques used to estimate 
$\Wln$ background contributions.  To minimize statistical 
uncertainties, the cross sections used to estimate backgrounds 
are a combination of the measurements made using both lepton 
decay channels. 

Estimates of the dijet $+$ \met candidate sample backgrounds 
from $Z$+jets production, in which the $Z$ boson decays to 
neutrinos, are taken directly from the measured $\Zllp$+jets 
cross sections based on the difference in the $Z$ branching 
fractions for charged leptons and neutrinos.  A second, 
independent estimate of this background is obtained from 
the measured $\Wlnp$+jets cross sections incorporating a 
theoretical prediction for $R_{(\frac{W}{Z})}$, the ratio 
of the $W$+jets and $Z$+jets production cross sections.  We
determine $R_{(\frac{W}{Z})}$ with a next-to-leading order 
(NLO) calculation using the {\sc mcfm} generator~\cite{mcfm}.
The value of $R_{(\frac{W}{Z})}$, which depends on the exact
choice of jet requirements, is calculated to be 8.7~$\pm$~0.2 
(8.2~$\pm$~0.2) for the loose (tight) dijet$+$\met sample.
The final background estimates are obtained by combining 
results from the two statistically-independent techniques 
which are found to be consistent.

Similarly, the measured $\Wlnp$+jets cross sections are 
used to extract $W$+jets background estimates for our 
dijet $+$ \met candidate samples.  The probability for 
the charged lepton in these events to fail the lepton 
veto criteria is obtained from the simulated event 
samples ($\sim$~20\% for electrons, $\sim$~33\% for muons, 
and $\sim$~55\% for taus) and applied as an acceptance 
factor on the measured cross section.  Smaller backgrounds 
from $Z$+jets, where the $Z$ boson decays into a pair of 
charged leptons that both fail lepton veto criteria, are 
estimated from the measured $\Zllp$+jets cross sections 
using the same technique.  Since the same measured cross 
sections are used to estimate all $W/Z$+jets backgrounds, 
the uncertainties on these predictions are fully 
correlated.  Small event contributions from $t\bar{t}$ 
and single-top production are obtained directly from 
simulated event samples.  We use a measured Run~II cross 
section~\cite{ttbar} for $t\bar{t}$ and a NLO cross 
section calculation~\cite{stop} for single-top production 
for the normalization of these samples.

The dominant multi-jet topology contributing events to our 
candidate samples is three-jet events in which the third jet is 
either not reconstructed or has an $E_T$ below our jet threshold 
(15~GeV).  The magnitude of this background is estimated from 
data using three-jet events in which the observed $\met$ points 
in the direction of the least-energetic jet.  We perform a 
linear extrapolation of the $E_{T}$ distribution obtained from 
the least-energetic jets in these events into the region where 
the $E_T$ falls below the threshold for defining jets.  Before 
performing the extrapolation, corrections obtained from 
simulation are applied to the distribution to remove $W/Z$+jets 
contributions.  A simulated \pyt~\cite{pythia} event sample is 
used to determine the relative fraction of events originating 
from other multi-jet topologies (20\%), and the result is used 
to scale the three-jet background estimates to account for 
all processes.  This scaling factor is assigned a conservative 
100\% uncertainty that does not contribute significantly to 
the total uncertainty on the multi-jet background estimate, 
which is dominated by statistical uncertainties due to the 
size of the three-jet candidate samples.  


The small background contribution from events in which a photon is 
produced in association with jets is taken from simulated samples 
generated with \pyt.  The estimates are normalized using a Run~II 
D0 measurement of the $\gamma$+jets cross section~\cite{d0gammajets}.  
The uncertainty associated with this measurement is the dominant 
contributor to the total uncertainty on the $\gamma$+jets background 
estimates.  Finally, the small, residual non-collision background is 
estimated using timing information from the hadronic calorimeter.


\begin{table}
\caption{Estimated SM backgrounds and the number of observed data
         events for loose ($H_T>$125~GeV, \met$>$80~GeV) and 
         tight ($H_T>$225~GeV, \met$>$100~GeV) candidate samples.}
\begin{center}
\begin{tabular}{lcc} \hline\hline
{Background}           & {Loose Sample}    & {Tight Sample}  \\\hline
{$\Znn$}               & { 888 $\pm$ 54}       & { 86.4 $\pm$ 12.7}  \\
{$\Wtn$}               & { 669 $\pm$ 42}       & { 50.6 $\pm$ 8.0}   \\
{$\Wmn$}               & { 399 $\pm$ 25}       & { 32.9 $\pm$ 5.2}   \\
{$\Wen$}               & { 256 $\pm$ 16}       & { 14.0 $\pm$ 2.2}   \\
{$\Zll$}               & {  29 $\pm$ 4}        & { 1.7 $\pm$ 0.2}    \\
{Top quark production} & {  74 $\pm$ 9}        & { 10.8 $\pm$ 1.7}   \\
{Multi-jet production} & {  49 $\pm$ 30}       & { 9.0 $\pm$ 9.0}    \\
{$\gamma$+jets}        & {  75 $\pm$ 11}       & { 4.8 $\pm$ 1.1}    \\
{Non-collision}        & {   4 $\pm$ 4}        & { 1.0 $\pm$ 1.0}    \\\hline
{Total expected}       & {2443 $\pm$ 151}      & { 211.2 $\pm$ 29.8} \\
{Data observed}        & {2506}                & {186}               \\\hline\hline
\end{tabular}
\end{center}
\label{allbgrs}
\end{table}

The estimated SM backgrounds and number of observed events in 
both the loose and tight dijet $+$ \met candidate samples are 
summarized in Table~\ref{allbgrs}.  The dominant source of 
uncertainty on the combined SM background predictions is the 
statistical size of the $\Wlnp$+jets and $\Zllp$+jets candidate 
samples (4.6\% and 12.2\% on the total background estimates for 
the loose and tight samples, respectively).  Other non-negligible 
uncertainty contributions come from the background estimates used 
in the $\Wlnp$+jets and $\Zllp$+jets cross section measurements 
(2.4\% and 4.0\%), the input parameters to the theoretical 
calculation of $R_{(\frac{W}{Z})}$ (1.8\% and 1.8\%), and the 
statistics of the three-jet samples used to perform the linear 
extrapolation for extracting multi-jet background estimates 
(1.2\% and 4.3\%).  The final combined uncertainties on the 
predicted SM backgrounds for the loose and tight candidate 
samples are 6.2\% and 14.1\%. 

In both the loose and tight candidate samples we observe no 
significant excess of events in data with respect to the SM
prediction, which constrains the potential contribution from 
new physics processes to these samples.  An upper limit on 
the number of non-SM signal events present in each candidate 
sample is obtained using a Bayesian approach with a flat 
prior for the number of signal events and priors based on 
gamma distributions for both the acceptance and the number 
of SM background events~\cite{statref}.  Limits on the 
number of signal events can be directly translated into 
upper limits on $\sigma \times A$ for any new physics 
process that contributes events to our candidate samples.  
These limits do not assume any central value for the 
signal acceptance, which is detector-dependent and varies 
significantly for different processes.  The quoted limits 
are based on a specific choice of values for acceptance 
uncertainties, which vary less among different processes.  
For a 15\% signal acceptance uncertainty we obtain a 
95~$\!\%$ C.L. upper limit of 0.18~pb (0.02~pb) on 
$\sigma \times A$ for the loose (tight) candidate sample.  
Increasing the signal acceptance uncertainty by a factor 
of two leads to a 25\% degradation in the quoted limits.

For the case of scalar leptoquark pair production with the 
subsequent decay of each leptoquark into a quark and neutrino, 
we provide an example of the detector-dependent acceptance 
calculation required to extract model limits.  We simulate 
signal acceptance using \pyt\ in conjunction with a full 
detector simulation.  The loose (tight) dijet$+$\met selection 
criteria yield an acceptance of 14\% (4\%) to a first generation 
leptoquark with a mass of 150~GeV/$c^2$.  Acceptance increases 
as a function of leptoquark mass ($M_{LQ}$), rising to 20\% (9\%) 
at 200~GeV/$c^2$.  The relative uncertainty on the acceptance 
is 13\% (20\%) independent of $M_{LQ}$ and comes from potential
variations in parton distribution functions (PDFs), ambiguity in 
the absolute jet energy scale~\cite{jetcor}, modeling of initial 
and final state radiation, data sample luminosity, and selection 
efficiencies.  Mass limits are based on a NLO production cross 
section calculation~\cite{kram} using the CTEQ6.1M PDF set~\cite{cteq} 
and $\mu = M_{LQ}$ for the renormalization and factorization scales.  
Uncertainties on the cross section due to PDF modeling (from the 
full set of CTEQ6.1M eigenvectors) and scale choice (from varying 
$\mu$ between $M_{LQ}/2$ and $2 \times M_{LQ}$) are added in 
quadrature.  We determine which sample has the best {\it a priori} 
sensitivity to the leptoquark model at each mass point, and set 
a 95\% C.L. lower mass limit based on the point where the cross 
section limit from the more sensitive sample intersects with the 
lower uncertainty band of the NLO calculation.  Figure~1 shows the 
cross section limits as a function of leptoquark mass that result 
in a lower mass limit of 187~GeV/$c^2$ for a first or second 
generation q$\nu$ scalar leptoquark (correponding to an upper 
cross section limit of 0.33~pb at this mass point).  This result 
significantly improves upon the previous CDF limit~\cite{cdflq} 
and is only slightly looser than the D0 lower mass limit of 
205~GeV/$c^2$~\cite{d0lq} obtained from an optimized search on 
a 25\% larger data sample.

\begin{figure}
\centering
\includegraphics[width=\linewidth]{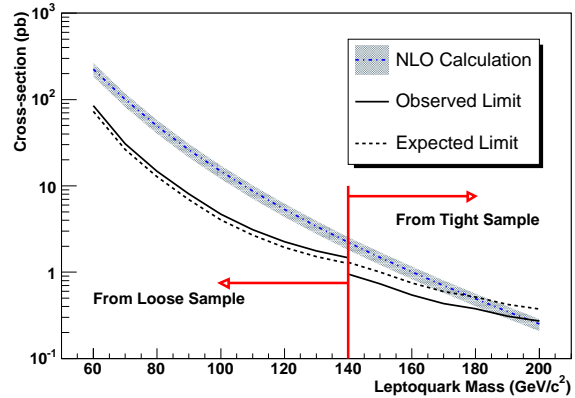}
\caption{95\% C.L. upper cross section limits on first and 
         second generation q$\nu$ scalar leptoquark pair 
         production (q being u,d,s or c) as a function of 
         leptoquark mass ($M_{LQ}$).}
\label{figlqlim}
\end{figure}

In summary, this article presents a signature-based 
search for potential non-SM contributions in the 
dijet$+$\met final state.  No excess above the SM 
prediction is observed and we set a 95\% C.L. upper 
limit on the cross section times acceptance for 
potential non-SM production processes.  For the 
specific case of first and second generation scalar 
leptoquark production, we obtain a 95\% C.L. lower 
mass limit of 187~GeV/$c^2$.

We thank M.~Kr\"{a}mer for calculating the next-to-leading
order leptoquark production cross sections. 
We thank the Fermilab staff and the technical staffs of 
the participating institutions for their vital contributions. 
This work was supported by the U.S. Department of Energy and 
National Science Foundation; the Italian Istituto Nazionale 
di Fisica Nucleare; the Ministry of Education, Culture, 
Sports, Science and Technology of Japan; the Natural Sciences 
and Engineering Research Council of Canada; the National 
Science Council of the Republic of China; the Swiss 
National Science Foundation; the A.P. Sloan Foundation; the 
Bundesministerium f\"ur Bildung und Forschung, Germany; 
the World Class University Program, the National Research 
Foundation of Korea; the Science and Technology Facilities 
Council and the Royal Society, UK; the Institut National 
de Physique Nucleaire et Physique des Particules/CNRS; the 
Russian Foundation for Basic Research; the Ministerio de 
Ciencia e Innovaci\'{o}n, and Programa Consolider-Ingenio 
2010, Spain; the Slovak R\&D Agency; and the Academy of 
Finland.


\begin{thebibliography}{0}
\expandafter\ifx\csname natexlab\endcsname\relax\def\natexlab#1{#1}\fi
\expandafter\ifx\csname bibnamefont\endcsname\relax
  \def\bibnamefont#1{#1}\fi
\expandafter\ifx\csname bibfnamefont\endcsname\relax
  \def\bibfnamefont#1{#1}\fi
\expandafter\ifx\csname citenamefont\endcsname\relax
  \def\citenamefont#1{#1}\fi
\expandafter\ifx\csname url\endcsname\relax
  \def\url#1{\texttt{#1}}\fi
\expandafter\ifx\csname urlprefix\endcsname\relax\def\urlprefix{URL }\fi
\providecommand{\bibinfo}[2]{#2}
\providecommand{\eprint}[2][]{\url{#2}}

\end{thebibliography}


\begin{thebibliography}{99}

\bibitem{metdef}
We use a coordinate system where $\theta$\ is the polar 
angle to the proton beam, $\phi$\ is the azimuthal angle 
about this beam axis, and $\eta$\ is the pseudorapidity 
defined as $-\ln\tan(\theta/2)$.  Missing transverse 
energy, \met, is defined as the magnitude of $-\sum_i
E_T^i \hat{n}_i$\, where $\hat{n}_i$\ is a unit vector 
in the azimuthal plane that points from the beamline to 
the $i$th calorimeter tower and $E_T^i$ is the transverse 
component of the measured energy in the tower, defined as 
$E^i \cdot \sin\theta$.

\bibitem{susy} G.L. Kane and J.P. Leville, Phys. Lett. B {\bf 112}, 227 (1982); 
               P.R. Harrison and C.H. Llewellyn-Smith, Nucl. Phys. {\bf B213}, 223 (1983).

\bibitem{ued} C. Macesanu, Int. J. Mod. Phys. A {\bf 21}, 2259-2296 (2006).

\bibitem{lq} J.L.~Hewett and T.G.~Rizzo, Phys. Rev. D {\bf 56}, 5709 (1997);
	     M.~Kr\"{a}mer, T. Plehn, M. Spira, and P. M. Zerwas, Phys. Rev. Lett. {\bf 79}, 341 (1997).

\bibitem{cdflq} D. Acosta {\it et al.} (CDF Collaboration), 
                Phys. Rev. D {\bf 72}, 051107 (2005);
		T. Aaltonen {\it et al.} (CDF Collaboration), 
                Phys. Rev. Lett. {\bf 102}, 121801 (2009).

\bibitem{d0lq} V.M.~Abazov {\it et al.} (D0~Collaboration), 
               Phys. Lett. B {\bf 668}, 357 (2008);
	       V.M.~Abazov {\it et al.} (D0~Collaboration), 
               Phys. Lett. B {\bf 660}, 449 (2008).

\bibitem{WZpaper} A. Abulencia {\it et al.} (CDF Collaboration),
                  J. Phys. G: Nucl. Part. Phys., 2457 (2007).

\bibitem{jetcor} A.~Bhatti {\it et al.}, Nucl. Instrum. Methods A {\bf 566}, 375 (2006).

\bibitem{ledprl} A. Abulencia {\it et al.} (CDF Collaboration), Phys. Rev. Lett. {\bf 97}, 171802 (2006).

\bibitem{alpgen} M.L. Mangano {\it et al.}, J. High Energy Phys. 07 (2003) 001.

\bibitem{geant} R.~Brun {\it et al.}, Tech. Rep. CERN-DD/EE/84-1 (1987);
                S.~Agostinelli {\it et al.}, Nucl. Instrum. Methods A {\bf 506}, 250 (2003).

\bibitem{mcfm} John~Campbell and R.K.~Ellis, Phys. Rev. D {\bf 65}, 113007 (2002).

\bibitem{ttbar} A. Abulencia {\it et al.} (CDF Collaboration), Phys. Rev. Lett. {\bf 97}, 082004 (2006).

\bibitem{stop} B.W. Harris {\it et al.}, Phys. Rev. D {\bf 66}, 054024 (2002); 
               Z. Sullivan, Phys. Rev. D {\bf 70}, 114012 (2004).

\bibitem{pythia} T.~Sj\"ostrand {\it et al.}, Comput. Phys. Commun. {\bf 135}, 238 (2001).

\bibitem{d0gammajets} V.M.~Abazov {\it et al.} (D0 Collaboration), Phys. Lett. B {\bf 639}, 151 (2006).

\bibitem{statref} J.~Heinrich {\it et al.}, arXiv:physics/0409129 (2004).

\bibitem{kram}  M.~Kr\"{a}mer {\it et al.}, Phys. Rev. Lett. {\bf 79}, 341 (1997).

\bibitem{cteq} J.~Pumplin {\it et al.}, J. High Energy Phys. 07 (2002) 012.
               D.~Stump {\it et al.}, J. High Energy Phys. 10 (2003) 046. 

\end{thebibliography}
\end{document}